\documentclass[12pt,showpacs,preprint,pra]{revtex4}
\usepackage{dcolumn,longtable}
\def\rhoo{\mbox{\boldmath{$\rho$}}}
\begin{document}

% J.-Y. Zhang, Z.-C. Yan, D. Vrinceanu, J. F. Babb, and H. R. Sadeghpour

\title{Long-range interactions for He($n\,S$)--He$(n'\,S)$ and He($n\,S$)--He$(n'\,P)$}
\author{J.-Y. Zhang$^1$, Z.-C. Yan$^1$, D. Vrinceanu$^{2}$, J. F. Babb$^3$, and H. R. Sadeghpour$^3$}

\affiliation{$^1$Department of Physics, University of New Brunswick, Fredericton, New Brunswick, Canada E3B 5A3}
\affiliation{$^2$Theoretical Division, Los Alamos National Laboratory, Los Alamos, NM 87545, USA}
\affiliation{$^3$ITAMP, Harvard-Smithsonian Center for
Astrophysics, Cambridge, Massachusetts 02138, USA}
\date{\today}

\begin{abstract}
The energetically lowest five states of a helium atom are: He($1^1S$), He($2^3S$),
He($2^1S$), He($2^3P$), and He($2^1P$). Long-range interaction coefficients $C_3$, $C_6$,
$C_8$, $C_9$, and $C_{10}$ for all $S-S$ and $S-P$ pairs of these states are calculated
precisely using correlated wave functions in Hylleraas coordinates. Finite nuclear isotope
mass effects are included.

\end{abstract} \pacs{34.20.Cf,31.15.Ar} \maketitle

In our previous paper~\cite{jzd02}, we presented precise
calculations of long-range interaction coefficients $C_3$, $C_6$,
$C_8$, $C_9$, and $C_{10}$ between two helium atoms in the
$2\,^3\!S$ and $2\,^3\!P$ states and the finite nuclear mass
effects for like isotopes. The purpose of this Brief Report is to
extend our calculations to include interactions between all $S-S$
and $S-P$ states of the energetically lowest five states:
He($1^1S$), He($2^3S$), He($2^1S$), He($2^3P$) and He($2^1P$).
These coefficients are useful in studying ultracold collisions
between two helium atoms~\cite{cohen}, as well as in serving as a
benchmark for other computational methods.

For two like isotope helium atoms $a$ and $b$, where one is in $S$ state and the other in
$L$ state with the associate  magnetic quantum number $M$, the zeroth-order wave function
for the combined system ab can be written in the form~\cite{yan96}:
\begin{eqnarray}
\Psi^{(0)} (M,\beta) &=& \frac{C}{\sqrt 2}[\Psi_{n_{a}}(\mbox{\boldmath{$\sigma$}})
\Psi_{n_b} (L M;\rhoo) +\beta \Psi_{n_{a}}(\rhoo)\Psi_{n_{b}}(L
M;\mbox{\boldmath{$\sigma$}})]\,, \label {eq:ap1}
\end{eqnarray}
where $\Psi_{n_{a}}$ is the $S$-state wave function, $\Psi_{n_{b}}$ is the $L$-state wave
function, $\mbox{\boldmath{$\sigma$}}$ and $\rhoo$ represent the coordinates of two helium
atoms, $C$ is the normalization factor, and $\beta$ describes the symmetry due to the
exchange of two atoms. If two atoms are both in a same $S$ state, $C$ is $\sqrt{2}$ and
$\beta$ is zero. If they are in different states, $C$ is 1 and $\beta$ is $\pm 1$.

At large internuclear distances $R$, the Coulombic interaction operator~\cite{yan96} between
two atoms can be expanded as an infinite series in powers of $1/R$
\begin{eqnarray}
V_{\rm op} &=& \sum_{\ell=0}^{\infty}\sum_{L=0}^{\infty} \frac{V_{\ell L}}{R^{\ell+L+1}}\,,
\label
{eq:ap2}
\end{eqnarray}
where
\begin{equation}
V_{\ell L}  = 4\pi (-1)^L (\ell,L)^{-1/2} \sum_{\mu} K_{\ell L}^{\mu} \;
T^{(\ell)}_{\mu}(\mbox{\boldmath{$\sigma$}})\; T^{(L)}_{-\mu}(\rhoo) \,. \label
{eq:ap3}
\end{equation}
$T^{(\ell)}_{\mu}(\mbox{\boldmath{$\sigma$}})$ and $T^{(L)}_{-\mu}(\rhoo)$ are the
atomic multipole tensor operators defined by
\begin{equation}
T^{(\ell)}_{\mu}(\mbox{\boldmath{$\sigma$}}) = \sum_{i} Q_i \sigma_i^\ell Y_{\ell
\mu}(\hat
{\mbox{\boldmath{$\sigma$}}}_i)\, ,\label{mpol}
\end{equation}
and
\begin{equation}
T^{(L)}_{-\mu}(\rhoo) = \sum_{j} q_j \rho_j^L Y_{L -\mu}(\hat {\rhoo}_j)\, ,
\end{equation}
where $Q_i$ and $q_j$ are the charges on particles $i$ and $j$. The coefficient
$K_{\ell L}^{\mu}$ is
\begin{eqnarray}
K_{\ell L}^{\mu} &=& \left[ { {\ell+L}\choose {\ell+\mu} } { {\ell+L}\choose {L+\mu}
}\right]^{1/2}\, \label {eq:ap4}
\end{eqnarray}
and $(\ell,L)=(2\ell+1)(2L+1)$.

For all $S-S$ systems and for $S-P$ systems with atoms in different spin symmetries, the
first order of perturbation theory vanishes and the second-order
perturbation in $V_{\rm op}$ gives rise to the following long-range interaction potential
\begin{eqnarray}
V &=&- \frac{C_{6}(M,\beta)}{R^{6}}- \frac{C_{8}(M,\beta)}{R^{8}}-
\frac{C_{10}(M,\beta)}{R^{10}}-\cdots\,, \label {eq:ap22}
\end{eqnarray}
where $C_{6}(M,\beta)$, $C_{8}(M,\beta)$, and $C_{10}(M,\beta)$ are the dispersion
coefficients. For resonant $S-P$ systems, where two atoms are in
a same spin state, the long-range interaction potential is
\begin{eqnarray}
V &=&- \frac{C_{3}(M,\beta)}{R^{3}}- \frac{C_{6}(M,\beta)}{R^{6}}-
\frac{C_{8}(M,\beta)}{R^{8}}-\frac{C_{9}(M,\beta)}{R^{9}}-
\frac{C_{10}(M,\beta)}{R^{10}}-\cdots\,, \label {eq:ap22_1}
\end{eqnarray}
where $C_{3}(M,\beta)$ is from the first-order energy correction,
$C_{6}(M,\beta)$, $C_{8}(M,\beta)$, and
$C_{10}(M,\beta)$ are from the second-order energy correction, and $C_{9}(M,\beta)$
is from the third-order energy correction~\cite{jzd02}.
The detailed expressions for $C_n$ and their evaluation in Hylleraas
coordinates can be found in \cite{jzd02}.

Tables~\ref{g1} and \ref{g2} present dispersion coefficients $C_{6}$, $C_{8}$, and $C_{10}$
for all six He($n\,^\lambda$S)--He($n'\,^{\lambda'}$S) systems with $n$ and $n' = 1,2$ and,
$\lambda$ and $\lambda' = 1,3$.
In Table~\ref{g1}, we also
include the revised values of the dispersion coefficients~\cite{yan96,yanbabb} for
He($1\,^1\!S$)--He($1\,^1\!S$), He($2\,^1\!S$)--He($2\,^1\!S$), and
He($2\,^3\!S$)--He($2\,^3\!S$) for the case of infinite nuclear mass, with some improvement,
particularly for $C_6$. Bishop and Pipin~\cite{bishop93} calculated the dispersion
coefficients for the system $^\infty\!\,$He($1\,^1\!S$)--$^\infty\!\,$He($2\,^3\!S$). Their
results in atomic units are $C_6=29.082914$, $C_8=1700.2700$, and $C_{10}=136380.30$, which
are in agreement with ours at the level of 1, 45, and 10 ppm respectively.

Table~\ref{g3} shows the contributions to $C_{6}(M,\beta)$ from different symmetries of
intermediate states for He($n\,^\lambda$S)--He($n'\,^{\lambda'}$P), except for
He($1\,^1\!S$)--He($2\,^3\!P$) and He($2\,^3\!S$)--He($2\,^3\!P$), which have been reported
in \cite{sadeg2} and \cite{jzd02} respectively. From this table, one can see that the
contributions from doubly-excited $(pp)P$ configurations are much smaller than the
contributions from singly-excited $S$ and $D$ configurations. It is also interesting to note
that only the contributions to $C_{6}({M,\pm})$ from the symmetries $(\,^1\!P,\,^3\!S)$ are
negative for $^\infty\!\,$He($2\,^1\!S$)--$^\infty\!\,$He($2\,^3\!P$). This is because the
dominant transitions $2\,^1\!S$--$2\,^1\!P$ and $2\,^3\!P$--$2\,^3\!S$ make large negative
contributions $-10935.32650$ and $-2733.831626$ for $C_{6}({0,\pm})$ and $C_{6}({\pm,\pm})$,
respectively. For the symmetries $(\,^3\!P,\,^1\!S)$ for
$^\infty\!\,$He($2\,^3\!S$)--$^\infty\!\,$He($2\,^1\!P$), the dominant transitions
$2\,^3\!S$--$2\,^3\!P$ and $2\,^1\!P$--$2\,^1\!S$ contribute $10935.32650$ and $2733.831626$
for $C_{6}({0,\pm})$ and $C_{6}({\pm,\pm})$, respectively. The mainly negative contributions
are from the pair of transitions $2\,^3\!S$--$2\,^3\!P$ and $2\,^1\!P$--$1\,^1\!S$ whose
values $-6.150732811$ and $-1.537683203$ for $C_{6}({0,\pm})$ and $C_{6}({\pm,\pm})$,
respectively, are much smaller than those from the corresponding dominant and positive
transitions.

Table~\ref{g4} lists
$C_{3}({M,\pm})$, $C_{6}({M,\pm})$, $C_{8}({M,\pm})$, $C_{9}({M,\pm})$, and
$C_{10}({M,\pm})$ for the He($n\,^\lambda$S)--He($n'\,^{\lambda'}$P) systems except for the
He($2\,^3\!S$)--He($2\,^3\!P$) system~\cite{jzd02}.
For $C_{3}({M,\pm})$, our results agree with Drake's values~\cite{drake}.
To our knowledge, there are no other published calculations on the dispersion coefficients
for the He($n\,^\lambda$S)--He($n'\,^{\lambda'}$P) system.

\acknowledgments
This work is supported by the Natural Sciences and Engineering Research Council of
Canada, by the ACRL of the University of New Brunswick, by the SHARCnet,
by the Westgrid, and by NSF through a
grant for the Institute of Theoretical Atomic, Molecular and Optical Physics (ITAMP) at
Harvard University and Smithsonian Astrophysical Observatory.

\newpage
\begin{longtable}{c c c c}
\caption{\label{g1} $C_{6}({0,\beta})$, $C_{8}({0,\beta})$, and $C_{10}({0,\beta})$, in
atomic units, for He($n\,^\lambda$S)--He($n'\,^{\lambda'}$S) ($n, n' = 1,2$ and $\lambda, \lambda' = 1,3$).
}\\
\hline\hline \multicolumn{1}{c}{System}& \multicolumn{1}{c}{$C_{6}({0,\beta})$}&
\multicolumn{1}{c}{$C_{8}({0,\beta})$}&
\multicolumn{1}{c}{$C_{10}({0,\beta})$}\\
\hline
$^\infty\!\,$He$(1\,^1\!
S)$--$^\infty\!\,$He$(1\,^1\!S)$&1.460977837725(2)&14.11785737(2)&183.691075(1)\\
$^4\!\,$He$(1\,^1\!
S)$--$^4\!\,$He$(1\,^1\!S)$&1.462122853192(3)&14.12578806(2)&183.781468(1)\\
%$^4\!\,$He$(1\,^1\!S)$--$^3\!\,$He$(1\,^1\!S)$&1.462310249395(2)&14.12708588(1)&183.796262(2)\\
$^3\!\,$He$(1\,^1\!
S)$--$^3\!\,$He$(1\,^1\!S)$&1.462497669977(2)&14.12838383(2)&183.811057(2)\\
$^\infty\!\,$He$(2\,^1\! S)$--$^\infty\!\,$He$(2\,^1\!S)$&11241.04684(4)&817250.26(2)
&108167575(3) \\
$^4\!\,$He$(2\,^1\!S)$--$^4\!\,$He$(2\,^1\!S)$&11247.73927(1)&817626.25(2)&108208732(3) \\
%$^4\!\,$He$(2\,^1\!S)$--$^3\!\,$He$(2\,^1\!S)$&11248.83446(3)&817687.76(2)&108215465(3) \\
$^3\!\,$He$(2\,^1\! S)$--$^3\!\,$He$(2\,^1\!S)$&
11249.92975(4)&817749.27(1)&108222197(1) \\
$^\infty\!\,$He$(2\,^3\! S)$--$^\infty\!\,$He$(2\,^3\!S)$&
3276.67964(5)&210566.54(3)&21786759(1)  \\
$^4\!\,$He$(2\,^3\!S)$--$^4\!\,$He$(2\,^3\!S)$&3279.45846(2)&210667.78(1)&21794920(2) \\
$^3\!\,$He$(2\,^3\!S)$--$^3\!\,$He$(2\,^3\!S)$&3280.36825(3)&210700.93(2)&21797593(3)  \\
$^\infty\!\,$He$(1\,^1\! S)$--$^\infty\!\,$He$(2\,^3\!S)$&29.082956(2)&1700.3495(4)&136381.56(2)\\
$^4\!\,$He$(1\,^1\!S)$--$^4\!\,$He$(2\,^3\!S)$&29.104446(2)&
1701.1618(1) &136444.57(3)\\
$^3\!\,$He$(1\,^1\!S)$--$^3\!\,$He$(2\,^3\!S)$&29.111482(3)&
1701.4278(1)&136465.17(1)\\
$^\infty\!\,$He$(2\,^1\!S)$--$^\infty\!\,$He$(2\,^3\!S)$&5817.46249(2)&417776.48(2)&49889993(3)\\
$^4\!\,$He$(2\,^1\!S)$--$^4\!\,$He$(2\,^3\!S)$&5821.97285(4)&417978.46(2)&49909416(2)\\
$^3\!\,$He$(2\,^1\!S)$--$^3\!\,$He$(2\,^3\!S)$&5823.44933(4)&418044.56(2) & 49915774(4)\\
\hline\hline
\end{longtable}

\begin{longtable}{c c c c}
\caption{\label{g2} $C_{6}({0,\pm})$, $C_{8}({0,\pm})$, and $C_{10}({0,\pm})$, in
atomic units, for He($1\,^1\!S$)--He($2\,^1\!S$).
}\\
\hline\hline \multicolumn{1}{c}{$C_n$}&
\multicolumn{1}{c}{$^\infty\!\,$He--$^\infty\!\,$He}&
\multicolumn{1}{c}{$^4\!\,$He--$^4\!\,$He}&
\multicolumn{1}{c}{$^3\!\,$He--$^3\!\,$He}\\
\hline $C_{6}({0,+})$&$44.750434(3)$&$44.783606(3)$&$44.794465(4)$ \\
$C_{6}({0,-})$&$38.932199(1)$&$38.961030(3)$& $38.970467(1)$\\
$C_{8}({0,+})$ &3406.535(3)&3408.196(2)&3408.739(1)\\
$C_{8}({0,-})$&$3214.354(3)$&$3215.913(2)$&$3216.425(3)$\\
$C_{10}({0,+})$&$353414.63(3)$&$353584.78(1)$& $353640.47(1)$\\
$C_{10}({0,-})$ &345572.365(3)&345738.985(3)&345793.517(3) \\
\hline\hline
\end{longtable}

\begin{longtable}{c c c}
\caption{\label{g3} Contributions to $C_{6}({M,\pm})$, in atomic units, for
$^\infty$He($n\,^\lambda$S)--$^\infty$He($n'\,^{\lambda'}$P) from different symmetries of
intermediate states.
}\\
\hline\hline
\multicolumn{1}{c}{$^\infty\!\,$He($1\,^1\!S$)--$^\infty\!\,$He($2\,^1\!P$)}&
\multicolumn{1}{c}{$C_{6}({0,\pm})$}&
\multicolumn{1}{c}{$C_{6}({\pm,\pm})$}\\
\hline
$(\,^1\!P,\,^1\!S)$ &29.80734050(1) &7.451835124(4)      \\
$(\,^1\!P,(pp)\,^1\!P)$ &0.054285059(2) &0.135712646(2)      \\
$(\,^1\!P,\,^1\!D)$ &29.011666(2) & 25.055529(1)      \\
\multicolumn{1}{c}{$^\infty\!\,$He($2\,^1\!S$)--$^\infty\!\,$He($2\,^1\!P$)}&
\multicolumn{1}{c}{$C_{6}({0,\pm})$}&
\multicolumn{1}{c}{$C_{6}({\pm,\pm})$}\\
\hline
$(\,^1\!P,\,^1\!S)$ &1296.69(1) &324.176(3)      \\
$(\,^1\!P,(pp)\,^1\!P)$ &1.237417(2) &3.093539(1)      \\
$(\,^1\!P,\,^1\!D)$ &4331.449(1) & 3740.798(1)      \\
\multicolumn{1}{c}{$^\infty\!\,$He($2\,^1\!S$)--$^\infty\!\,$He($2\,^3\!P$)}&
\multicolumn{1}{c}{$C_{6}({0,\pm})$}&
\multicolumn{1}{c}{$C_{6}({\pm,\pm})$}\\
\hline
$(\,^1\!P,\,^3\!S)$ &$-9439.938\,18(1)$ &$-2359.984\,54(2)$     \\
$(\,^1\!P,(pp)\,^3\!P)$ &1.880\,576\,3(3) &4.701\,440\,3(3)    \\
$(\,^1\!P,\,^3\!D)$ &3199.141\,6(2) &2762.894\,89(1)    \\
\multicolumn{1}{c}{$^\infty\!\,$He($2\,^3\!S$)--$^\infty\!\,$He($2\,^1\!P$)}&
\multicolumn{1}{c}{$C_{6}({0,\pm})$}&
\multicolumn{1}{c}{$C_{6}({\pm,\pm})$}\\
\hline
$(\,^3\!P,\,^1\!S)$ & 11520.93(2)&2880.22(1)   \\
$(\,^3\!P,(pp)\,^1\!P)$&0.8674943(1)&2.1687357(2)   \\
$(\,^3\!P,\,^1\!D)$ &2602.9197(1)&2247.9762(2)   \\
\hline\hline
\end{longtable}

\begin{longtable}{c c c c}
\caption{\label{g4}  $C_{3}({M,\pm})$, $C_{6}({M,\pm})$, $C_{8}({M,\pm})$,
$C_{9}({M,\pm})$,
and $C_{10}({M,\pm})$, in atomic units, for He($n\,^\lambda$S)--He($n'\,^{\lambda'}$P).
}\\
\hline\hline \multicolumn{1}{c}{$C_n$}&
\multicolumn{1}{c}{$^\infty\!\,$He($1\,^1\!S$)--$^\infty\!\,$He($2\,^1\!P$)}&
\multicolumn{1}{c}{$^4\!\,$He($1\,^1\!S$)--$^4\!\,$He($2\,^1\!P$)}&
\multicolumn{1}{c}{$^3\!\,$He($1\,^1\!S$)--$^3\!\,$He($2\,^1\!P$)}\\
\hline
  $C_{3}({0,\pm})$&$\pm0.3541112056(4)$&$\pm0.3541530610(1)$&$\pm0.3541667599(1)$\\
  $C_{3}({\pm,\pm})$&$\mp0.1770556028(2)$&$\mp0.1770765307(3)$&
$\mp0.17708337999(6)$\\
$C_{6}({0,\pm})$&58.873293(3)&58.923259(2)&58.939615(1)\\
 $C_{6}({\pm,\pm})$&32.643077(1)&32.670985(3)&32.680118(1)\\
 $C_{8}({0,+})$&9693.3127(2)&9700.5496(3)&9702.9182(3)\\
$C_{8}({0,-})$&10048.0818(1)&10055.4826(3)&10057.9047(2)\\
 $C_{8}({\pm ,+})$&396.5492(2)&396.4958(2)&396.4782(2)\\
$C_{8}({\pm ,-})$&357.4777(2)&357.4055(3)&357.3817(2)\\

$C_{9}({0,\pm})$&$\mp271.24449(2)$&$\mp271.55655(3)$&$\mp271.65869(1)$  \\
$C_{9}({\pm,\pm})$&$\pm76.76195(2)$&$\pm76.85062(2)$ &$\pm76.87965(2)$\\
$C_{10}({0,+})$&1235282.3(3)&1236209.5(3)&1236512.8(1)\\
$C_{10}({0,-})$&1252541.6(3)&1253475.7(2)&1253781.5(2)\\
$C_{10}({\pm ,+})$&20803.39(2)&20815.53(3)&20819.48(1)\\
$C_{10}({\pm ,-})$&19268.36(2)&19279.86(3)&19283.62(2) \\
\multicolumn{1}{l}{}&
\multicolumn{1}{c}{$^\infty\!\,$He($1\,^1\!S$)--$^\infty\!\,$He($2\,^3\!P$)}&
\multicolumn{1}{c}{$^4\!\,$He($1\,^1\!S$)--$^4\!\,$He($2\,^3\!P$)}&
\multicolumn{1}{c}{$^3\!\,$He($1\,^1\!S$)--$^3\!\,$He($2\,^3\!P$)}\\
\hline
$C_{6}({0,\pm})$&47.725\,886\,76(2)&47.752\,349\,75(3)&47.761\,010\,19(2)  \\
$C_{6}({\pm,\pm})$&26.708\,670\,89(1)&26.723\,515\,42(2) &26.728\,373\,55(2)\\
$C_{8}({0,\pm})$ &7129.97(1)&7131.59(1)&7132.13(2)\\
$C_{8}({\pm,\pm})$&281.55(2)& 281.38(1)& 281.33(2)  \\
$C_{10}({0,\pm})$&801678.7(1)& 801717.2(3)& 801729.6(2)\\
$C_{10}({\pm,\pm})$ &13105.8(1)& 13104.6(2)& 13103.9(1) \\
\multicolumn{1}{l}{}&
\multicolumn{1}{c}{$^\infty\!\,$He($2\,^1\!S$)--$^\infty\!\,$He($2\,^1\!P$)}&
\multicolumn{1}{c}{$^4\!\,$He($2\,^1\!S$)--$^4\!\,$He($2\,^1\!P$)}&
\multicolumn{1}{c}{$^3\!\,$He($2\,^1\!S$)--$^3\!\,$He($2\,^1\!P$)}\\
\hline
  $C_{3}({0,\pm})$&$\pm17.0096686055(3)$&$\pm17.0157415780(8)$&$\pm17.0177287170(5)$\\
  $C_{3}({\pm ,\pm})$&$\mp8.5048343028(2)$&$\mp8.5078707890(4)$& $\mp8.5088643585(3)$\\
$C_{6}({0,\pm})$&5629.39(1)&5634.35(5)&5635.95(4)\\
 $C_{6}({\pm ,\pm})$&4068.07(1)&4071.68(1)&4072.87(2)\\
 $C_{8}({0,+})$&679008(3)&679615(3) &679815(4)\\
$C_{8}({0,-})$&4600179(4)&4603085(5)&4604034(3)\\
 $C_{8}({\pm ,+})$&492965(4)&493226(3)&493312(3)\\
$C_{8}({\pm ,-})$&419817(3)&419939(1)&419985(4)\\
$C_{9}({0,\pm})$&$\pm1719978(5)$&$\pm1722188(5)$&$\pm1722914(3)$  \\
$C_{9}({\pm,\pm})$&$\mp366261.1(5)$&$\mp366698.1(2)$ &$\mp366841.3(4)$\\
$C_{10}({0,+})$&$6.27699(1)\times 10^{7}$&$6.28316(2)\times 10^{7}$&$6.28518(2)\times
10^{7}$\\
$C_{10}({0,-})$&$6.031831(1)\times 10^{8}$ &$6.035509(1)\times
10^{8}$&$6.036714(2)\times 10^{8}$ \\
$C_{10}({\pm ,+})$&$5.791649(1)\times 10^{7}$&$5.794589(3)\times
10^{7}$&$5.795549(2)\times 10^{7}$\\
$C_{10}({\pm ,-})$&$1.119933(3)\times 10^{7}$&$1.120226(3)\times
10^{7}$&$1.120319(1)\times 10^{7}$ \\
\multicolumn{1}{l}{}&
\multicolumn{1}{l}{$^\infty\!\,$He($2\,^1\!S$)--$^\infty\!\,$He($2\,^3\!P$)}&
\multicolumn{1}{l}{$^4\!\,$He($2\,^1\!S$)--$^4\!\,$He($2\,^3\!P$)}&
\multicolumn{1}{l}{$^3\!\,$He($2\,^1\!S$)--$^3\!\,$He($2\,^3\!P$)}\\
\hline
$C_{6}({0,\pm})$& $-6238.9163(2)$&$-6253.30080(2)$&$-6258.01373(1)$\\
$C_{6}({\pm,\pm})$&407.611780(1)&404.83595(2)&403.92612(2)\\
$C_{8}({0,\pm})$ &1962689(1)&1962880(1)&1962943(1)\\
$C_{8}({\pm,\pm})$&389203.6(2)&389217.5(3)&389221.9(1)\\
$C_{10}({0,\pm})$&$2.2430065(3)\times
10^{8}$&$2.2429754(5)\times10^{8}$&$2.2429647(2)\times 10^{8}$\\
$C_{10}({\pm,\pm})$ &$2.828207(2)\times 10^{7}$ &$2.828419(1)\times 10^{7}$
&$2.828489(2)\times 10^{7}$ \\
\multicolumn{1}{l}{}&
\multicolumn{1}{l}{$^\infty\!\,$He($2\,^3\!S$)--$^\infty\!\,$He($2\,^1\!P$)}&
\multicolumn{1}{l}{$^4\!\,$He($2\,^3\!S$)--$^4\!\,$He($2\,^1\!P$)}&
\multicolumn{1}{l}{$^3\!\,$He($2\,^3\!S$)--$^3\!\,$He($2\,^1\!P$)}\\
\hline $C_{6}({0,\pm})$&14124.71(2) &14143.81(1)&14150.06(2)  \\
$C_{6}({\pm,\pm})$&5130.37(1)&5136.64(2)&5138.71(1)\\
$C_{8}({0,\pm})$ &1288203(1)&1289227(1)&1289563(1)\\
$C_{8}({\pm,\pm})$&162321(1)& 162400(1)& 162424(2)  \\
$C_{10}({0,\pm})$&$1.640216(1)\times 10^{8}$& $1.6413357(2)\times
10^{8}$&
$1.6417021(2)\times 10^{\,8}$  \\
$C_{10}({\pm,\pm})$ &$1.161798(1)\times 10^{7}$& $1.162275(2)\times 10^{7}$&
$1.162431(2)\times 10^{7}$\\
\hline\hline
\end{longtable}


\begin{thebibliography}{99}
\bibitem{jzd02} J.-Y. Zhang, Z.-C. Yan, D. Vrinceanu, J. F. Babb, and H. R. Sadeghpour,
Phys.\ Rev.\ A {\bf 73}, 022710 (2006).
\bibitem{cohen} J. L\'{e}onard, M. Walhout, A. P. Mosk,
T. M\"{u}ller, M. Leduc, and C. Cohen-Tannoudji, Phys.\ Rev.\ Lett.\ {\bf 91}, 073203
(2003).
\bibitem{yan96} Z.-C. Yan, J. F. Babb, A. Dalgarno, and G. W. F. Drake,
Phys.\ Rev.\ A {\bf 54}, 2824 (1996).
\bibitem{yanbabb} Z.-C. Yan and J. F. Babb, Phys.\ Rev.\ A {\bf 58}, 1247 (1998).
\bibitem{bishop93} D. M. Bishop and J. Pipin, Int.\  J.\  Quantum\ Chem.\ {\bf 47}, 129
(1993); {\it ibid.} {\bf 45}, 349 (1993).
\bibitem{sadeg2} J.-Y. Zhang, Z.-C. Yan, D. Vrinceanu, and H. R. Sadeghpour,  Phys.\
Rev.\ A {\bf 71}, 032712 (2005).
\bibitem{drake} G. W. F. Drake, {\it Atomic, Molecular, and Optical Physics Handbook},
edited by G. W. F. Drake (AIP woodbury, NY, 1996), p. 169.
\end{thebibliography}
\end{document}